\documentclass[a4paper,fleqn]{cas-sc}

\usepackage[sort&compress,numbers]{natbib}

\usepackage{xcolor}

\begin{document}
\let\WriteBookmarks\relax
\def\floatpagepagefraction{1}
\def\textpagefraction{.001}

\shortauthors{A. Alekseev and K. J. Kapcia}
%
%
\title [mode = title]{Charge-ordered states and the phase diagram of
the extended Hubbard model on the Bethe lattice}

\shorttitle{Charge-ordered states and the phase diagram of
the extended Hubbard model on the Bethe lattice}

\author[1]{Aleksey Alekseev}[%
                    orcid=0000-0001-5102-6647]
\ead{aleksey.alekseev@amu.edu.pl}
\credit{Software, Formal analysis, Investigation, Data curation, Writing - Original draft preparation, Writing - Review \& Editing, Visualization}

\author[1]{Konrad Jerzy Kapcia}[%
                    orcid=0000-0001-8842-1886]
\ead{konrad.kapcia@amu.edu.pl}
\credit{Conceptualization, Methodology, Validation, Investigation, Resources, Writing - Original draft preparation, Writing - Review \& Editing, Supervision, Project administration, Funding acquisition}

\address[1]{Institute of Spintronics and Quantum Information, Faculty of Physics and Astronomy, Adam Mickiewicz University in Pozna\'n, Uniwersytetu Pozna\'nskiego 2, PL-61614 Pozna\'n, Poland}

\begin{abstract}
We study the extended Hubbard model (EHM) with both onsite Hubbard interaction and the intersite density-density interaction between nearest-neighbors using the standard Hartree mean-field approximation (MFA) on the Bethe lattice.
We found that, at the ground state, the system can be in a charge-ordered band-insulating (COI), a charge-order metallic (COM) or a non-charge-ordered (NO) state.
Moreover, the finite-temperature phase diagrams are presented. 
Several observables like a charge-order parameter, a spectral function, and particularly at finite temperatures, a thermally-excited charge carrier concentration (to visualize the degree of metallicity) are analyzed.  
The results show that increasing onsite repulsion suppresses charge order and change the properties of the system from insulating to metallic.
Worth noting, that a number of phenomena can be found within the MFA, where their analysis is much simpler than in more advanced approaches. The method used for the EHM on the Bethe lattice also allows for a series of analytical derivations and simplifications to see general geometry-independent features and analytical results, avoiding the numerical inaccuracies and other issues that appear with a purely numerical solution.
\end{abstract}

%
%
%
%
\begin{keywords}
extended Hubbard model \sep Bethe lattice \sep charge order \sep metal-insulator transition \sep mean-field approximation \sep phase diagram 
\end{keywords}
%
%
%
\maketitle
%
%
%

\section{Introduction}

Charge-order phenomena have been widely observed in various real compounds and have attracted broad interest not least due to its interplay and competition with superconductivity \cite{MRR1990,NeupertNP2022,KangNM2023}.
For instance they have been evidenced experimentally in hole-doped perovskite-type transition-metal oxides (cuprates, manganites, bismutates), transition-metal dichalcogenides \cite{MorosanNP2006,NovelloPRB2015}, numerous organic compounds, \cite{HirakiPRL1998,NodaPRL2002}, moir\'{e} systems \cite{kumar2025}, heavy-fermion systems, and various systems with alternating (mixed) valence. These phenomena encompasses Wigner crystals, including their generalized version, charge-density waves, charge-transfer insulators \cite{Kuramoto1978,Matty2022,Tan2023}.

The model for investigation of the charge ordering is the extended Hubbard model (EHM) \cite{RobaszkiewiczPSSB1973,RobaszkiewiczAPPA1974,HirschPRB1984,HirschPRB1985,LinPRB1986}, which is, despite its simplicity, the most natural improvement of even more simple and artificial Hubbard model \cite{Hubbard1963,PennPRB1966,ChaoPRB1979,SpalekPRB1987} that provided a lot of valuable insights into strongly-correlated physics. In this paper we study the EHM with onsite and intersite (nearest-neighbor) density-density interactions without attachment to a particular class of materials. We consider such studies of a model itself an advantageous step towards understanding of possible charge-order phenomena and mechanisms of their formation, as well as hints where to look for unusual phenomena. Worth noting, we investigate the model in the grand canonical ensemble with an all-encompassing range of chemical potentials without limiting to particular filling factors. To our knowledge, such an approach is often ignored despite the fact that the fixed-occupation approach can result in incorrect phase diagram when the chosen occupation appears within phase-separated states. The phase separation correspond to the case where two phases with discontinuous transition between each other macroscopically coexist to form the state with intermediate spatially-averaged order parameter that lies within the range of the discontinuous jump of this parameter.

The model has been investigated within the atomic limit (without hopping term) \cite{MRC1984,BursillJPCM1993,BJK1996,FRU2001,MM2008,KKR2010,Mancini2010,KR2011,KR2011b,KKR2012,KapciaPhysA2016}, including also the Monte Carlo method \cite{P2008,PK2008,PG2008}. 
The investigation within the broken symmetry Hartree-Fock approximation with finite hopping amplitude can be found in  \cite{RobaszkiewiczPSSB1973,RobaszkiewiczAPPA1974,AlekseevPRB2025} (particularly in the context of the high-temperature superconductors (cuprates) \cite{MRR1990}). 
The spinless model corresponds to a model without the Hubbard (onsite interaction) term and has been investigated in \cite{VlamingJPCM1992,UV1993a,UhrigJPCM1993,ShankarRMP1994,UV1995,KK2001,ZTYT2005,FTZ2006,CRT2008,CGNR2012a,CGNR2012b}; the research so far concentrates on the case of half-filled band with nearest-neighbor hopping. 
There are dynamical mean-field theory (DMFT) studies for charge density $n=1/2$ at zero temperature \cite{ACH2010} and finite temperature \cite{PBB1999}; for arbitrary charge density, onsite interaction $U=2D$ and small intersite interaction $V = 1.2D,1.4D$ ($D$ is a half-bandwidth) on the Bethe lattice \cite{TSB2004}. The all-encompassing ground-state research within the DMFT for the same model as in this work can be found in Ref. \cite{kapcia2017}.
Finally, there are the cellular DMFT results at $n=1/2$ and zero temperature on a square lattice \cite{M2007}; $GW$+DMFT results at $n=1$ on a square lattice \cite{AWB2012,ABW2013}; and extended DMFT and $GW$+EDMFT results on square and cubic lattices for longer-range intersite interactions, $n=1$ and close to it \cite{HAB2014}. The results with other extensions of DMFT have been also obtained \cite{TerletskaPRB2021,IskakovPRB2022,PhiloxenePRB2024,KunduSciPost2024}.

Here, we restrict ourselves to the case of commensurate checkerboard ordering on the alternate lattices and use the mean-field (broken-symmetry Hartree) approximation (MFA) with semicircular (Bethe) noninteracting density of states to solve the model. Despite existing ground-state results within DMFT for the same model \cite{kapcia2017}, our research serves several purposes. First, our intention is to fill the gap in the literature to clearly track the correlation-induced effects when comparing it with the DMFT result, since previous MFA studies of the EHM were focused on half-filled phases only \cite{RobaszkiewiczPSSB1973,RobaszkiewiczAPPA1974,HirschPRB1984,LinPRB1986} and consider other lattices or non-interacting densities of states (DOS) (constant DOS \cite{RobaszkiewiczPSSB1973,RobaszkiewiczAPPA1974}, 1D lattice \cite{HirschPRB1984,LinPRB1986}, triangular lattice \cite{AlekseevPRB2025}). Worth noting, that a number of phenomena can be found within the MFA where their analysis is much simpler. We also consider finite temperatures in contrast to \cite{kapcia2017}. Second, the mean-field solution of the EHM on the Bethe lattice allows for a series of analytical derivations and simplifications. We consider it as a toy model to see general geometry-independent features and analytical results, while the paper itself can serve pedagogical purposes.

The work is organized as follows. In Sec. \ref{sec:model}, the model and the approach are introduced and the general expression are derived. Sec. \ref{sec:spectral} is devoted for important observables, which are useful for the phase identification. 
Sec. \ref{sec:gs} provides the results at the ground state (Sec. \ref{sec:gs-simexp} for analytical expressions, whereas Sec. \ref{sec:gs-results} for results obtained from them). In Sec. \ref{sec:ft} the results for finite temperatures are presented. Finally, Sec. \ref{sec:conclusions} contains conclusions and final remarks.

\section{Model and method}\label{sec:model}

In this work, we investigate the extended Hubbard model \cite{MRR1990,PBB1999,kapcia2017} in the form of
\begin{equation}\label{eq:EHM}
    \hat{H} 
    = - t \sum_{(i,j)\sigma} \hat{c}^\dag_{i\sigma} \hat{c}_{j\sigma}
    + U\sum_i \hat{n}_{i\uparrow}\hat{n}_{i\downarrow}
    + \frac{V}{2} \sum_{(i,j)} \hat{n}_i \hat{n}_j
    - \mu\hat{N},
\end{equation}
where $t$ is a hopping amplitude, $U$ and $V$ are density-density onsite and nearest-neighbor (NN) intersite interaction parameters, $\mu$ is a chemical potential, $\sum_{(i,j)}$ means a summation over all lattice sites $i$ and $j$ that are nearest neighbors to each other, $\sigma$ is a spin index ($\uparrow$, $\downarrow$), $\hat{c}^\dag_{i\sigma}$ and $\hat{c}_{i\sigma}$ are creation and annihilation operators, $\hat{n}_{i\sigma} = \hat{c}^\dag_{i\sigma}\hat{c}_{i\sigma}$ is an occupation number operator, $\hat{n}_i = \sum_\sigma \hat{n}_{i\sigma}$, and $\hat{N} = \sum_i \hat{n}_i$.

We consider the checkerboard orderings or equivalently a 2-sublattice assumption, where each NN pair consists of two lattice sites from different sublattices (sublattice index is $\alpha=A,B$).
After the Fourier transform to a reciprocal space and a mean-field decoupling, neglecting exchange (Fock) and pairing (anomalous, superconducting) parts of the decoupling (i.e., $\hat{n}_{i\sigma}\hat{n}_{j\sigma'} \overset{\text{MF}}{=} n_{j\sigma'} \hat{n}_{i\sigma} + n_{i\sigma} \hat{n}_{j\sigma'} - n_{i\sigma} n_{j\sigma'}$, where $n_{i\sigma} \equiv \langle \hat{n}_{i\sigma} \rangle$), the Hamiltonian (\ref{eq:EHM}) turns into a sum of independent terms:
\begin{equation}\label{eq:ham}
    \hat{H}
    = \sum_{\mathbf{k}\sigma} \left[ 
    \varepsilon_\mathbf{k} \hat{c}^\dag_{A\mathbf{k}\sigma} \hat{c}_{B\mathbf{k}\sigma} + 
    \varepsilon_\mathbf{k}^* \hat{c}^\dag_{B\mathbf{k}\sigma} \hat{c}_{A\mathbf{k}\sigma} 
    + E_{A\sigma} \hat{n}_{A\mathbf{k}\sigma}
    + E_{B\sigma} \hat{n}_{B\mathbf{k}\sigma} \right] + C.
\end{equation}
It is also called the broken-symmetry Hartree-Fock approach, although without the Fock-part.
Here, $\mathbf{k}$ is a reciprocal-space vector, 
$\varepsilon_\mathbf{k}$ are Fourier-transform components of the hopping term, 
$\sum_\mathbf{k}$ means a summation over $\frac{L}{2}$ vectors in the (reduced) first Brillouin zone, $L$ is the number of lattice sites enclosed by the periodic boundary conditions,
\begin{equation}
    C = -\frac{L}{2} \left[ U n_{A\uparrow}n_{A\downarrow} + U n_{B\uparrow}n_{B\downarrow} + zV n_A n_B \right]
\end{equation}
is the constant term,
\begin{equation}
    E_{\alpha\sigma}
    = U n_{\alpha\bar\sigma} + zV n_{\bar\alpha} - \mu,
\end{equation}
$n_{\alpha\sigma} = \frac{2}{L} \sum_{i\in\alpha} n_{i\sigma} = \frac{2}{L} \sum_\mathbf{k} n_{\alpha\mathbf{k}\sigma}$,
$n_\alpha = \sum_\sigma n_{\alpha\sigma}$,
$z$ is the coordination number, 
and $\bar\sigma$ or $\bar\alpha$ denotes a spin or sublattice index, different from $\sigma$ or $\alpha$, respectively,

In this research, in correspondence to previous works, we focus on the charge-ordering phenomenon neglecting possible magnetic orderings. 
Thus, we take $n_{\alpha\sigma} = \frac{n_\alpha}{2}$, and hence $E_{\alpha\uparrow} = E_{\alpha\downarrow} \equiv E_\alpha$.

We use the total electron density ($n$) and charge polarization (charge-order parameter, $\Delta$) as order parameters, defined as
\begin{align}
    n = \frac{1}{2}\left(n_A + n_B\right),
    &&
    \Delta = \frac{1}{2}\left(n_A - n_B\right).
\end{align}

Let us introduce the quantities
\begin{align}\label{eq:E0-EAB}
    E_0 = \frac{E_A + E_B}{2}
    = \left(zV + \frac{U}{2}\right)n - \mu,
    &&
    E_{AB} = \frac{E_B - E_A}{2} = 
    \left(zV - \frac{U}{2}\right)\Delta,
\end{align}
which are useful in the following discussion, and their physical meaning is discussed in Sec. \ref{sec:spectral}.
An analytical diagonalization of the Hamiltonian (\ref{eq:ham}) gives the following self-consistent system of equations for the order parameters:
\begin{eqnarray}\label{eq:sceA}
    n & = &\frac{2}{L} \sum_\mathbf{k}
    [f_\text{FD}(\epsilon_\mathbf{k}^-)
    + f_\text{FD}(\epsilon_\mathbf{k}^+)],
    \\
    \label{eq:sceB}
    \Delta & = &\frac{2}{L} \sum_\mathbf{k}
    \frac{E_{AB}}{Q_\mathbf{k}}
    [f_\text{FD}(\epsilon_\mathbf{k}^-)
    - f_\text{FD}(\epsilon_\mathbf{k}^+)],
\end{eqnarray}
where
\begin{equation}
    f_\text{FD}(\epsilon) = \frac{1}{1 + e^{\beta\epsilon}},
\end{equation}
is the Fermi-Dirac distribution, $\beta=T^{-1}$, $T$ is a temperature (for $\beta=\infty$, $f_\text{FD}(\epsilon)=\theta(-\epsilon)$, where $\theta$ is the Heaviside step function), 
\begin{align}
    \epsilon_\mathbf{k}^\pm = E_0 \pm Q_\mathbf{k}
    ,&&
    Q_\mathbf{k} = \sqrt{E_{AB}^2 + |\varepsilon_\mathbf{k}|^2}.
\end{align}

To investigate the model (\ref{eq:EHM}) we use a semicircular density of state (DOS) which is the noninteracting DOS of the Bethe lattice with $z\rightarrow\infty$:
\begin{equation}\label{eq:NonIntDos}
    \rho(\varepsilon) = \frac{2}{L} \sum_\mathbf{k} \delta(\varepsilon - \varepsilon_\mathbf{k})
    =\frac{2}{\pi D} \sqrt{1 - \left( \frac{\varepsilon}{D} \right)^2},\quad \text{for } |\varepsilon|<D
\end{equation}
(per spin), where $D=2t$. 
Note that it is not that different from some cubic-lattice DOSs \cite{GeorgesRMP1996}.
The corresponding noninteracting half-bandwidth as well as a typical kinetic energy is $2\sqrt{\int^{\infty}_{-\infty}\varepsilon^2\rho(\varepsilon)d\varepsilon} = D$.
We express the quantities, such as $U$, $zV$, $\mu$, and $T$, in the units of $D$ (or equivalently, we set $D=1$).
We also use the expression 
\begin{equation}
    \frac{2}{L} \sum_\mathbf{k} f(\varepsilon_\mathbf{k})
    = \int \rho(\varepsilon) f(\varepsilon) d\varepsilon
\end{equation}
to solve  self-consistent system (\ref{eq:sceA})--(\ref{eq:sceB}). 
A strict convergence criterion of $10^{-12}$ is set for the order parameters. 

To find which phase is stable in coexistence regions (for fixed $\mu$) we compare their grand potentials expressed by
\begin{equation}
    \Omega = C - \frac{2}{\beta} \sum_\mathbf{k} \ln \mathcal{Z}_\mathbf{k},
\end{equation}
where
\begin{equation}\label{eq:grandpot}
    \mathcal{Z}_\mathbf{k} = 1 + e^{-\beta\epsilon_\mathbf{k}^-} 
    + e^{-\beta\epsilon_\mathbf{k}^+}
    + e^{-\beta(\epsilon_\mathbf{k}^- + \epsilon_\mathbf{k}^+)}.
\end{equation}
For the ground state ($T=0D$) it is also equal to the total energy ($E=\langle\hat{H}\rangle$) and takes a simplified form of
\begin{equation}
    \Omega = E = C + 2\sum_\mathbf{k} \min\{0, E_0 - Q_\mathbf{k}, 2E_0 \}.
\end{equation}

In the research, we make use of the shifted chemical potential defined as $\bar\mu = \mu - zV - \frac{U}{2}$, chosen that way that $\bar\mu=0D$ corresponds to the half-filling of the lattice ($n=1$) and it is also a point of a particle-hole symmetry, i.e., the results for $\bar\mu < 0D$ ($n<1$) can be easily mapped to the results for $\bar\mu > 0D$ ($n>1$).

The calculations for solving eqs. (\ref{eq:sceA})--(\ref{eq:sceB}) are implemented in Python language, and the figures are made using its \texttt{matplotlib} library.

\section{Spectral function and thermally-excited charge carriers}\label{sec:spectral}

\begin{figure}
    \centering
    \includegraphics[width=0.55\textwidth]{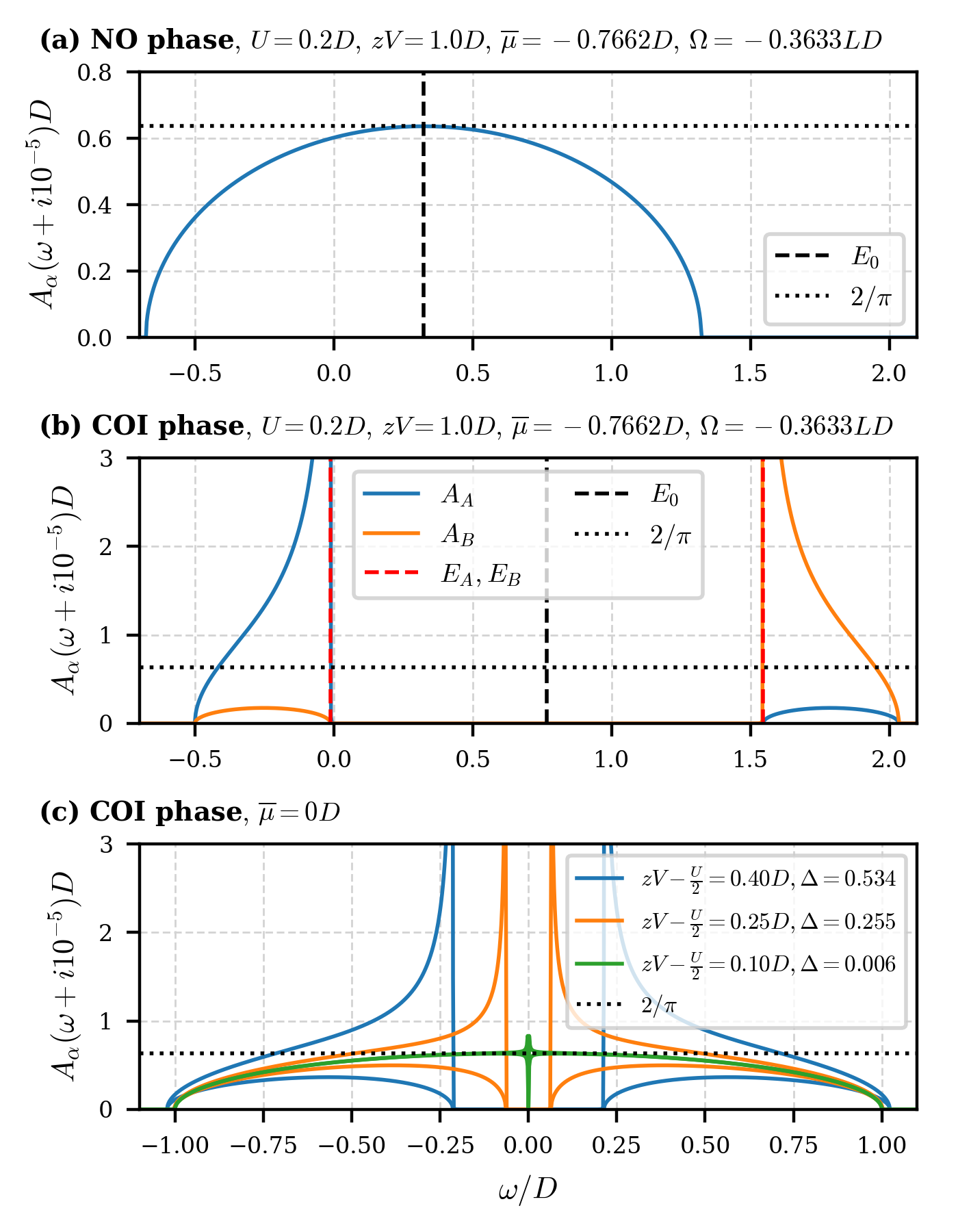}
    \caption{Spectral functions for the ground state. (a) and (b) are spectral functions of the non-charge-ordered (NO, $n=0.597$, $\Delta=0$) metal and charge-ordered band insulator (COI, $n=1$, $\Delta=0.865$) for the same point of the phase diagram where their grand potentials ($\Omega$) are equal (the point on a discontinuous phase transition line). (c) is a spectral function evolution of the COI phase at $\bar\mu=0D$ when $zV - \frac{U}{2}$ approaches $0D$ (the continuous transition to the half-filled NO phase).}
    \label{fig:spectral}
\end{figure}

An additional tool to interpret the ground-state results is a spectral function ($A_\alpha(\omega + i\eta)$), while its value at the Fermi level ($\omega=0D$) is another order parameter (the $\lim_{\eta\rightarrow+0} A_\alpha(i\eta)$ is zero for band-insulating phases and nonzero for metallic phases, which can be ordered or non-ordered). The spectral function ($\alpha$-sublattice projected) is calculated as
\begin{equation}
    A_\alpha(\omega + i\eta) = -\frac{2}{\pi}\frac{2}{L}\sum_\mathbf{k} \operatorname{Im} G_{\alpha\alpha\mathbf{k}\sigma}(\omega + i\eta)
\end{equation}
and the total spectral function is defined as
\begin{equation}
    A(\omega + i\eta) = \sum_\alpha A_\alpha(\omega + i\eta),
\end{equation}
with $\eta = 10^{-4}D$,
\begin{equation}
    G_{\alpha\alpha\mathbf{k}\sigma}(z)
    = \frac{\xi_{\bar\alpha}(z)}{\xi_A(z) \xi_B(z) - |\varepsilon_\mathbf{k}|^2},
\end{equation}
and $\xi_\alpha(z) = z - E_\alpha$. 
With the semicircular noninteracting DOS (i.e., for the Bethe lattice) one gets:
\begin{equation}
    \frac{2}{L}\sum_\mathbf{k}G_{\alpha\alpha\mathbf{k}\sigma} = \frac{2}{L}\sum_\mathbf{k} \frac{\xi_{\bar\alpha}}{\xi_A \xi_B - |\varepsilon_\mathbf{k}|^2}
    =\begin{cases}
        \frac{\xi_{\bar\alpha}}{D^2} \left(1 - i\frac{\sqrt{D^2 - \xi_A\xi_B}}{\sqrt{\xi_A\xi_B}} \right), &\operatorname{if}\quad\operatorname{Im}[\xi_A\xi_B] \ge 0, \\
        \frac{\xi_{\bar\alpha}}{D^2} \left(1 + i\frac{\sqrt{D^2 - \xi_A\xi_B}}{\sqrt{\xi_A\xi_B}} \right), &\operatorname{if}\quad\operatorname{Im}[\xi_A\xi_B] < 0.
    \end{cases}
\end{equation}

Figs. \ref{fig:spectral}a and \ref{fig:spectral}b show typical shapes of the spectral function at $T=0D$ for a non-charge-ordered (NO) phase with $\Delta = 0$ and a charge-ordered (CO) phase with $|\Delta|>0$, respectively. Both of them are calculated for the same point of a ground-state phase diagram, where these two phases have the same total energy (the point of a discontinuous phase transition). 
For the NO phase, the $E_0=E_A=E_B$, $E_{AB}=0D$, and the spectral function is simply a semicircular DOS shifted by the value of $E_0$. 
At $\omega=E_0$ the spectral function is maximal and equal to $2/\pi$ (per sublattice). 
Note that the self-consistent solution of the system (\ref{eq:sceA})--(\ref{eq:sceB}) for the NO phase always exists within the mean-field approximation. 
At $T=0D$, for the $|\bar\mu| \ge zV + \frac{U}{2} + D$, the NO phase is a fully-unoccupied or fully-occupied band insulator.

The spectral function of the CO phase shows two bands: the lower one primarily comes from a more occupied sublattice, and the upper one---from a less occupied sublattice. In the large-$V$ limit, the $n_\alpha$ are equal to $2$ and $0$, and there is a full correspondence between a sublattice and a band. At the ground state, the position of the Fermi level inside or outside of a band defines if the phase is a charge-ordered band insulator (COI) or a charge-ordered metal (COM), respectively. 
The COI phase is always half-filled (i.e., $n_\text{COI}=1$).

If the hopping is suppressed (atomic limit), the CO phase consists of two atomic levels at the points $E_A$ and $E_B$ with a gap $2|E_{AB}|=|E_B-E_A|$ and a middle point $E_0$ between them. With the finite hopping, the levels becomes finite-width bands. Usefully, the points $E_A$ and $E_B$ correspond to singularities at the very edges of the bands, so that the value of $2|E_{AB}|$ is a band gap at the ground state. Note that there are severe convergence problems in the self-consistent system (\ref{eq:sceA})--(\ref{eq:sceB}) for the COM phase when the Fermi level is close to the singularities (for $E_\alpha \approx 0D$).

At finite temperatures ($T>0D$), the order parameter $A_\alpha(i\eta)$ does not provide useful information due to the thermal excitations beyond the Fermi level. 
Instead, to visualize the degree of metallicity or thermally-excited charge carrier concentration, we use the quantity defined as
\begin{equation}\label{eq:c}
    c = \int_{-\infty}^{0} A(\omega+i\eta) f_\text{FD}(-\omega) d\omega 
    + \int_{0}^{\infty} A(\omega+i\eta) f_\text{FD}(\omega) d\omega.
\end{equation}
Its physical meaning is a total concentration of thermally-excited conduction electrons and holes when such a division of charge carriers is still applicable.
Note that this quantity is zero at $T=0D$ even for metallic phases, but it grows fast as $T$ increases.

\section{Ground-state results}\label{sec:gs}

\subsection{Simplified expressions at $T=0$}\label{sec:gs-simexp}

A number of simplifications of the self-consistent equations (\ref{eq:sceA})--(\ref{eq:sceB}) are possible for the $T=0D$ (ground state) case. 
Below we present analytical formulas in this limit.

\subsubsection{NO phase}
For the NO phase, the order parameter $\Delta=0$ and the self-consistent equations reduce to
\begin{equation}\label{eq:nNO}
    n_\text{NO} = \frac{2}{\pi} \arccos\frac{E_0}{D} - E_0 \rho(E_0)
\end{equation}
with $E_0$ from eq. (\ref{eq:E0-EAB}).
The expression for the total energy is also simplified to
\begin{equation}
    \frac{E_\text{NO}}{L} = \frac{C}{L} +
    n_\text{NO} E_0 -\frac{\pi^2 D^4}{6} \rho(E_0)^3,
\end{equation}
i.e., it is analytical as far as $n_\text{NO}$ is found numerically from eq. (\ref{eq:nNO}).

\subsubsection{COI phase}\label{sec:COI}

Let us introduce a function
\begin{equation}
        \mathcal{E}_\text{h}(x | m) = -i\mathcal{E}(i\operatorname{arcsinh}x \,|\, m)
\end{equation}
where $\mathcal{E}(\varphi|m)$ is an incomplete elliptic integral of the second kind. Note that the function does not actually contains imaginary units: $-i\mathcal{E}(i\varphi|m) = \int_0^{\varphi} \sqrt{1 + m\sinh^2\theta} \,d\theta$. 
We also use notations: $\mathcal{F}(\varphi|m)$ for the incomplete elliptic integral of the first kind ($\mathcal{F}(\varphi|m) = \int_0^\varphi (1 - m \sin^2 \theta )^{-1/2}\, d \theta $), and $\mathcal{K}(m) = \mathcal{F}(\frac{\pi}{2}|m)$ and $\mathcal{E}(m) = \mathcal{E}(\frac{\pi}{2}|m)$ for the complete elliptic integrals.

For $\Delta \neq 0$ and $|E_0| \le |E_{AB}|$, the system is always at the half-filling (i.e., $n=1$) and the charge polarization becomes
\begin{eqnarray}\label{eq:DeltaCOI}
    \Delta_\text{COI} 
    &= &\frac{4|E_{AB}|}{\pi D} \mathcal{E}_\text{h}\left(\frac{D}{E_{AB}} \left| -\! \left(\frac{E_{AB}}{D}\right)^2\right.\right)
    \\
    \nonumber
    &=&
    \frac{4|E_{AB}|}{\pi D} \frac{E_{AB}}{D}
    \left[ \left(q^{-1} + 1 \right) \mathcal{K}(-q^{-1}) - \mathcal{E}(-q^{-1}) \right],
\end{eqnarray}
where $q = \left( E_{AB}/D \right)^2$. 
The opposite is also correct: for $n=1$ and $\Delta \ne 0$, the $|E_0|$ (which is also equal to the $|\bar\mu|$) is always less or equal to the $|E_{AB}|$. Thus, eq. (\ref{eq:DeltaCOI}) is a self-consistent equation for the $\Delta$ of a half-filled CO phase, more specifically, of the COI phase. It does not require to perform numerical integration and numerically much more stable than the original equation. 
Notably, $\Delta_\text{COI}$ does not depend on $\mu$, as well as points $E_A$, $E_B$, and the distance $E_{AB}$ on a spectral function plot. Moreover, only the difference $(zV-\frac{U}{2})$ matters.

The condition $|E_{AB}| = |\bar\mu|$ defines points where the COI phase continuously changes into the COM phase (the point of a continuous transition, if the CO phases are stable with respect to the NO phase at this point). 
In the large $(zV - \frac{U}{2})$ limit, the $\Delta_\text{COI}$ asymptotically approaches $1$, and an asymptotic line of the continuous transition is determined as $(zV - \frac{U}{2}) = |\bar\mu|$.

A band gap of the COI phase is found as
\begin{equation}
    E_\text{gap} = 2|E_{AB}| = 2\left(zV - \frac{U}{2}\right)|\Delta_\text{COI}|.
\end{equation}

\begin{figure}
    \centering
    \includegraphics[width=0.99\textwidth]{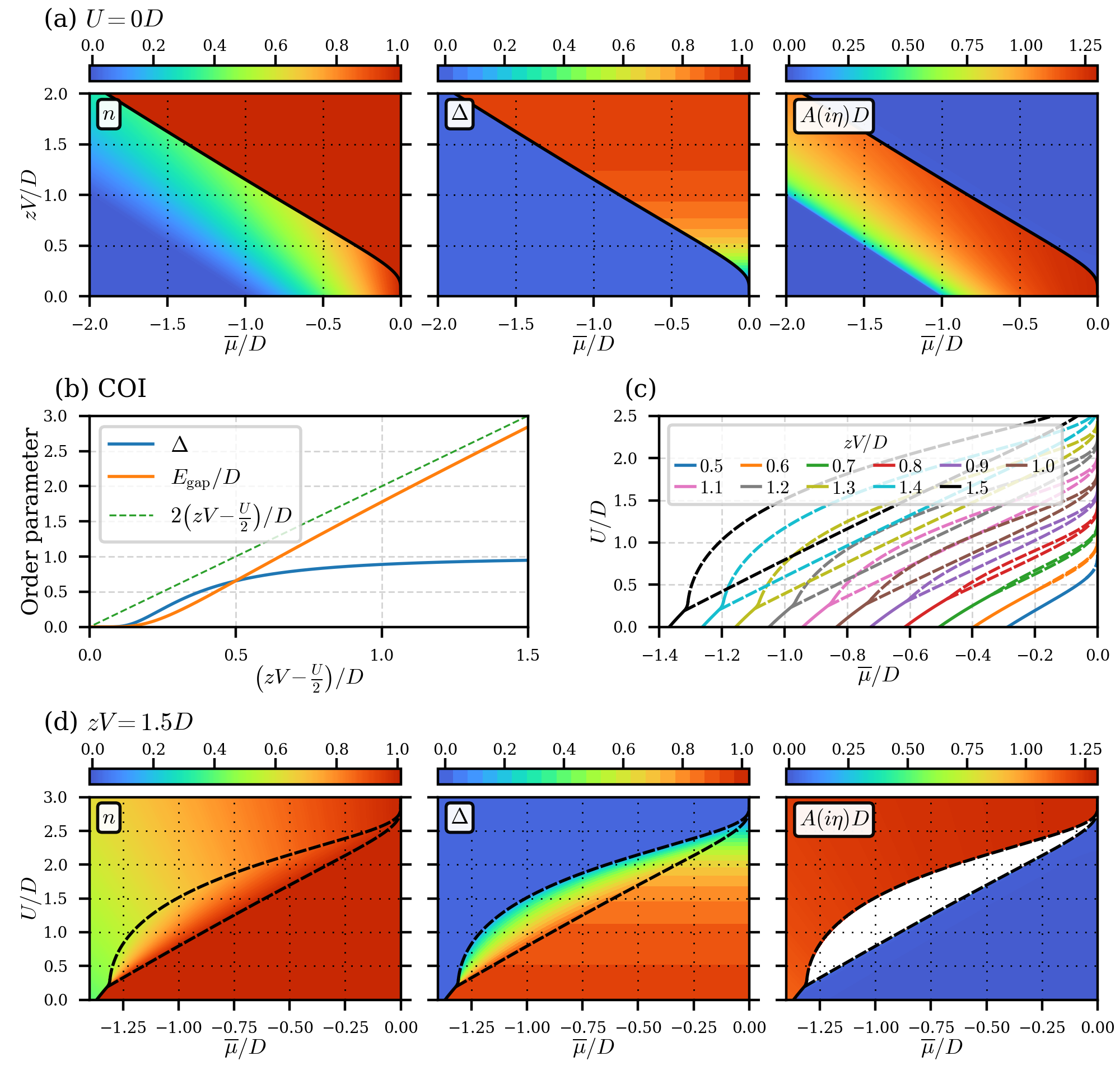}
    \caption{Ground-state phase diagrams and properties. 
    (a)  $\bar\mu$-$V$ phase diagram for $U=0D$; 
    (b) dependencies of COI-phase properties; 
    (c) evolution of the $\bar\mu$-$U$ phase diagram when $V$ changes; 
    (d) $\bar\mu$-$U$ phase diagram for $V=1.5D$. 
    On the phase diagrams: dashed and solid black lines denote continuous and discontinuous transitions, respectively;
    color scales are for concentration $n$, charge polarization $\Delta$ and spectral function $A(i\eta)D$, $\eta=10^{-4}$ (as labeled).
    The white region on the $A(i\eta)D$ color-contour plot corresponds to the model parameters, where the value of $A(i\eta)D$ is far out of a range of the presented colorbar (it is unrepresentative to extend the colorbar range), the corresponding phase is the COM phase. 
    Only the phases with the lowest grand potential are shown (no metastable phases).}
    \label{fig:GS}
\end{figure}

Both quantities, $\Delta_\text{COI}$ and $E_\text{gap}$, are shown as a function of $\left(zV-\frac{U}{2}\right)$ in Fig. \ref{fig:GS}b.

The total energy of the COI phase is also simplified so that the numerical integration is not required.
Note that it helps to avoid numerical inaccuracies when the $\left(zV - \frac{U}{2}\right)$ is very small.
The expression for the energy is
\begin{equation}
    \frac{E_\text{COI}}{L} = \frac{C}{L} -\bar\mu
    - \frac{4|E_{AB}| }{3\pi}
    \left[ (1 + q) \mathcal{K}(-q^{-1}) + (1 - q) \mathcal{E}(-q^{-1}) \right],
\end{equation}
i.e., its only dependency on the chemical potential is the term $-\bar\mu$.

It is clear, that the second equation in system (\ref{eq:sceA})--(\ref{eq:sceB}) cannot yield $\Delta \ne 0$ (a CO phase) for $zV < \frac{U}{2}$. 
In fact, at $\bar\mu=0D$ there is a continuous transition from the half-filled COI phase to the half-filled NO phase (COI-NO transition) exactly at $zV = \frac{U}{2}$. As $(zV - \frac{U}{2})$ approaches zero, from above, one gets
\begin{equation}
    \Delta_\text{COI} \approx \pm \frac{4D}{zV - \frac{U}{2}} \exp\left[-\frac{\pi D}{4 \left(zV - \frac{U}{2}\right)} - 1\right].
\end{equation}
The corresponding spectral functions of the COI phase are shown in Fig. \ref{fig:spectral}c. The asymptotic behaviour of the $\Delta$ and the $(E_\text{COI}-E_\text{NO})$ around this continuous transition makes the COI and NO phases indistinguishable up to a machine precision ($\sim 10^{-16}$ for $(E_\text{COI}-E_\text{NO})/(LD)$) for a noticeable region of parameters ($zV-\frac{U}{2} \lesssim 0.05D$, see also Fig. \ref{fig:GS}b), while the original equations, that involve numerical integration, yield incorrect results for a larger region of $(zV-\frac{U}{2})$ (around $0.1D$). For this reason, the correct results in this region can be obtained with the presented analytical simplifications only.

\subsubsection{COM phase}

When $|E_0| > |E_{AB}|$ and $\Delta \ne 0$, the set of equations describes the COM phase. 
The system of self-consistency equations reduces to
\begin{eqnarray}\label{eq:sceCOMA}
    n_\text{COM} &= &1 - \operatorname{sign}(E_0) \left[1 - n_\text{c}\left(-\sqrt{E_0^2 - E_{AB}^2}\right) \right]
    =\begin{cases}
        n_\text{c}\left(-\sqrt{E_0^2 - E_{AB}^2}\right), & \text{for}\ E_0 > 0D, \\
        2 - n_\text{c}\left(-\sqrt{E_0^2 - E_{AB}^2}\right), & \text{for}\ E_0 < 0D,
    \end{cases}
    \\
    \label{eq:sceCOMB}
    \Delta & = &
    \Delta_\text{COI} - \frac{4|E_{AB}|}{\pi D}
    \mathcal{E}_\text{h}\left(\frac{\sqrt{E_0^2 - E_{AB}^2}}{E_{AB}} \left| -\! \left(\frac{E_{AB}}{D}\right)^2\right.\right),
\end{eqnarray}
where we have introduced the following function
\begin{equation}
    n_\text{c}(\mu) = \frac{2}{\pi} \arccos\left( \frac{-\mu}{D} \right) + \mu \rho(\mu),
\end{equation}
which is a cumulative charge density when the noninteracting semicircular DOS is integrated and $\rho(\varepsilon)$ is defined by (\ref{eq:NonIntDos}). 
Particularly, eq. (\ref{eq:nNO}) can be written as $n_\text{NO} = n_\text{c}(-E_0) = 2 - n_\text{c}(E_0)$.

The total energy for the COM phase is also simplified to the expression
\begin{eqnarray}
    \frac{E_\text{COM}}{L} & = & \frac{C}{L} + E_\text{COI}
    + \frac{2E_0}{3} \sqrt{E_0^2 - E_{AB}^2} \cdot \rho\left(\sqrt{E_0^2 - E_{AB}^2}\right)
    \\ 
    \nonumber
    & + & \frac{4|E_{AB}| }{3\pi}
    \left[ (1 + q) \mathcal{F}\left(\left.\arcsin\frac{\sqrt{E_0^2 - E_{AB}^2}}{D} \,\right|\, \!-\! q^{-1} \right)
    + (1 - q) \mathcal{E}\left(\left.\arcsin\frac{\sqrt{E_0^2 - E_{AB}^2}}{D} \,\right|\, \!-\! q^{-1}\right) \right],
\end{eqnarray}
so that it does not require numerical integration as far as $n_\text{COM}$ and $\Delta$ are found.

The continuous transition COM-NO (when it exists) for specific $zV/D$ and $U/D$, can be easily located solving the following system of equations
\begin{eqnarray}\label{eq:cont-COM-NO-A}
    n & = & n_\text{c}(-E_0),
    \\
    \label{eq:cont-COM-NO-B}
    1 & = & \frac{4}{\pi D}\left(zV-\frac{U}{2}\right)
    \left(\ln\left[ 1 + \frac{1}{D}\sqrt{D^2 - E_0^2} \right]
    - \ln \left( \frac{|E_0|}{D} \right) 
    -\frac{1}{D}\sqrt{D^2 - E_0^2}
    \right).
\end{eqnarray}
with respect to $n$ and $\bar\mu$. The direct solution of system (\ref{eq:sceA})--(\ref{eq:sceB}) or simplified system (\ref{eq:sceCOMA})--(\ref{eq:sceCOMB}) at the point of the continuous COM-NO transition requires extremely large (millions) number of iterations during which $\Delta$ slowly converges to $0$. 
This makes using the 
system (\ref{eq:cont-COM-NO-A})--(\ref{eq:cont-COM-NO-B}) advantageous. 
Moreover, usefully, when this system gives a solution at which $\Delta \ne 0$, it indicates, that the continuous transition COM-NO does not exist, and the transition is discontinuous only (at some other point).

\begin{figure}
    \centering
    \includegraphics[width=0.9\textwidth]{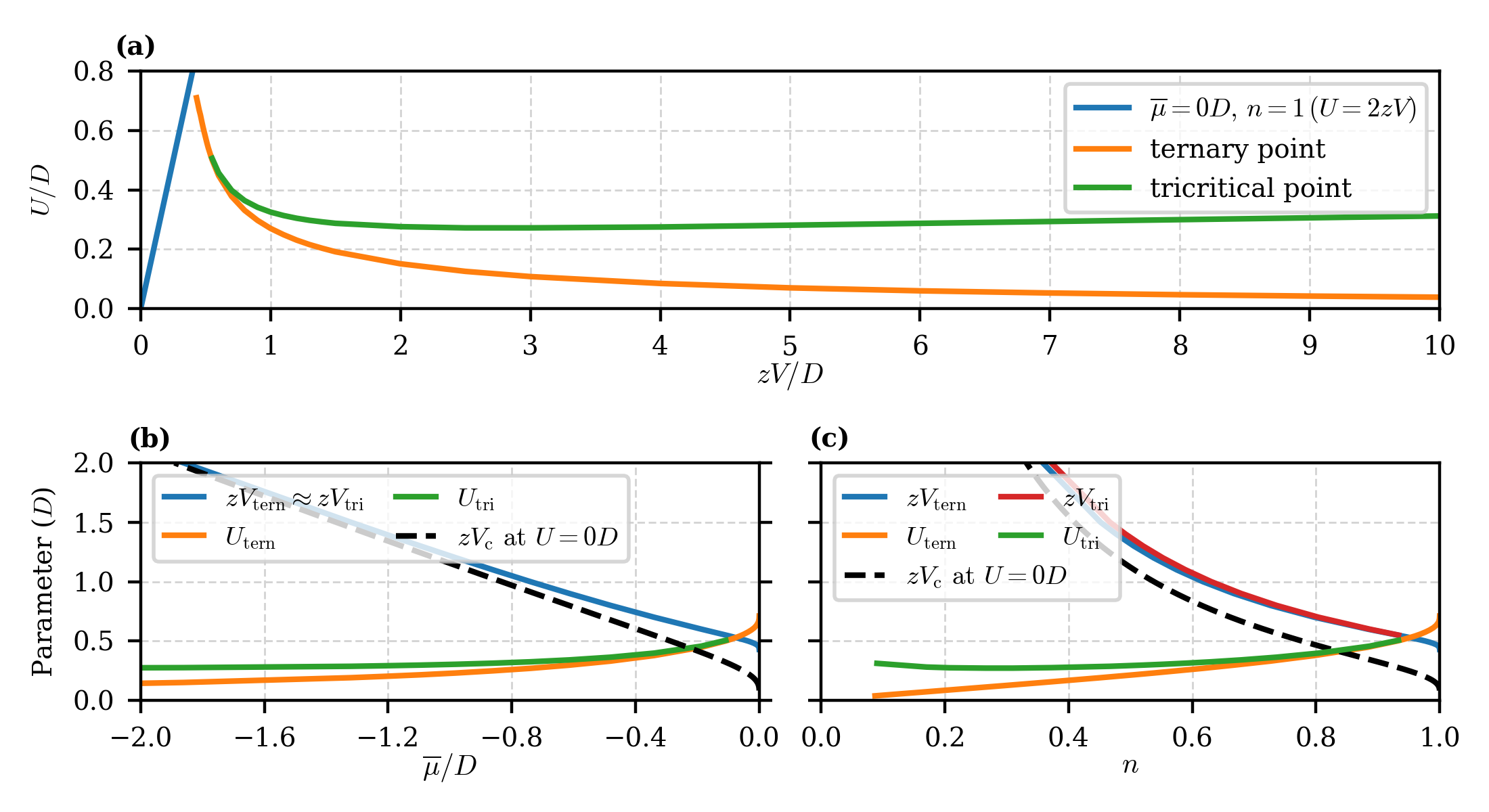}
    \caption{Tendencies of the ground-state critical points: a relation between the interaction strengths of the critical points, alongside with a point of the continuous COI-NO transition at $\bar\mu=0D$ (a); and dependencies of the interaction strengths of the critical points on the shifted chemical potential (b) and the electron density of the NO phase (c), alongside with a point of the discontinuous COI-NO transition at $U=0D$ ($V_\text{c}$, Fig. \ref{fig:GS}a).}
    \label{fig:GS-critical}
\end{figure}

\begin{figure}
    \centering
    \includegraphics[width=0.99\textwidth]{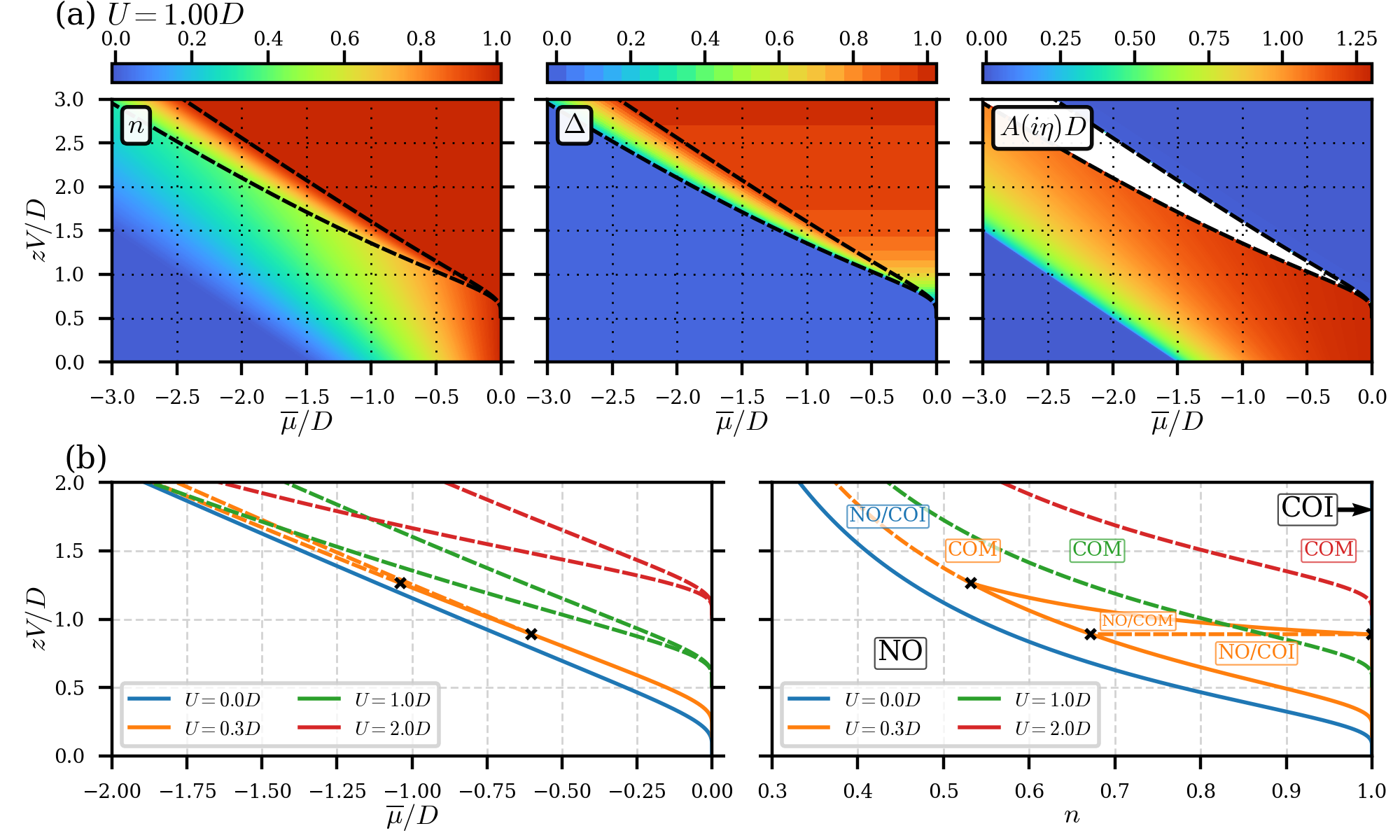}
    \caption{Ground-state phase diagrams and properties. 
    (a)  $\bar\mu$-$V$ phase diagram for $U=1.00D$; 
    (b) evolution of the $\bar\mu$-$V$ (left) and $n$-$V$ (right) phase diagrams when $U$ changes with indicated phase-separated states on the $n$-$V$ phase diagrams.
    Black crosses in (b) correspond to the ternary and the tricritical points for $U=0.3D$. Other denotations on the phase diagrams are as in Fig. \ref{fig:GS}.}
    \label{fig:GS-fixU}
\end{figure}

\subsection{Results at $T=0$}\label{sec:gs-results}

Fig. \ref{fig:GS}a shows the $U=0D$ ground-state phase diagram at hole doping for a range of repulsive intersite interactions. Hereinafter, only the phases with the lowest grand potential $\Omega$ are shown on phase diagrams. 
The COI phase is identified as a phase with nonzero charge polarization ($\Delta \ne 0$), half-filling ($n=1$), and the lack of the spectral weight at the Fermi level ($A(i\eta)D=0$). As the chemical potential decreases, the COI phase discontinuously transitions to the NO metallic phase slightly before it would continuously transition to the COM phase (i.e., before $\bar\mu$ riches value $-\left(zV-\frac{U}{2}\right)|\Delta|$). At $\bar\mu = 0D$ the transition to the half-filled NO phase is continuous as follows from Fig. \ref{fig:GS}b. As mentioned before, the NO phase becomes a fully-unoccupied insulator for the $\bar\mu \le -(zV+\frac{U}{2}+D)$.

When the onsite repulsion is nonzero and the chemical potential decreases, the COI phase can continuously transition to the COM phase (with $0 < \Delta < 1$, $n_\text{NO} < n < 1$, $A(i\eta)D > 0$) before it gets unstable toward the NO phase (the example for $zV=1.5D$ is shown in Fig. \ref{fig:GS}d). 
Thus, there is a point where the COI, NO, and COM phases meet, which is called a ternary point ($V_\text{tern}$, $U_\text{tern}$, $\bar\mu_\text{tern}$). 
In the studied system, at the ternary point (which is a higher-order critical point \cite{GriffithsPRB1975}) two discontinuous boundaries (COI-NO, COM-NO) and continuous boundary (COI-COM) merge.
For $U$ slightly larger than $U_\text{tern}$, the COM-NO transition is discontinuous similarly to the located nearby COI-NO transition.
Further, however, the transition becomes continuous ($\Delta \rightarrow 0$, $n_\text{COM} \rightarrow n_\text{NO}$ at the boundary), and thus, a tricritical point ($V_\text{tri}$, $U_\text{tri}$, $\bar\mu_\text{tri}$) is identified (associated with a change of a type of the COM-NO transition). One can see the evolution of the $\bar\mu$-$U$ phase diagram for varying $V$ in Fig. \ref{fig:GS}c. Additionally, Fig. \ref{fig:GS-critical} shows the evolution of the ternary and tricritical points from different perspectives.
The horizontal axis in Fig. \ref{fig:GS-critical}c is the electron density of the NO phase ($n_\text{NO}$), because at the ternary point the $n_\text{COI}$ and $n_\text{COM}$ always equal $1$, and at the tricritical point $n_\text{COM}=n_\text{NO}$. For a fixed $V$, $\bar\mu_\text{tern} \approx \bar\mu_\text{tri}$, and vice versa, for a fixed $\bar\mu$, $V_\text{tern} \approx V_\text{tri}$.

Fig. \ref{fig:GS-fixU}a and the left panel of Fig. \ref{fig:GS-fixU}b show phase diagrams from the perspective of fixed onsite interaction $U$ which is used to identify strong-correlation effects when comparing with the DMFT results from \cite{kapcia2017} (see also Sec. \ref{sec:conclusions}). Since the electron density experiences a jump at the point of discontinuous transition from $n_{\text{NO}}$ to $n_{\text{CO}}$ ($n_{\text{COM}}$ or $n_{\text{COI}}$), a range of $n$ can manifest itself within the macroscopic phase separation only. As shown in \cite{ArrigoniPRB1991,BakAPPA2004,KR2011,KR2011b,KapciaPhysA2016}, the equal value of grand potentials of two phases at $\bar\mu$ of discontinuous transition, ensures the global stability of phase-separated states with $n$ in the range of $n_{\text{NO}}<n<n_{\text{CO}}$, with $n_\text{NO}$ and $n_\text{CO}$ that can be found from the Maxwell's construction (and equal to those found at the first order boundary for fixed $\bar{\mu}$). 
The right panel of Fig. \ref{fig:GS-fixU}b shows the phase-separated states that are formed as a result of the discontinuous transitions NO-COI and NO-COM at $U=0.0D$ and $0.3D$ (regions NO/COI and NO/COM; the latter is only for $U=0.3D$).
The range of repulsive interaction parameters where the phase separation takes place is everything below the line of tricritical points in Fig. \ref{fig:GS-critical}a. As the intersite interaction increases, the ternary point asymptotically approaches the $U=0D$, while the $U$ of the tricritical point slightly increases, hence the range of $U$ with the phase separation increases as well.

For small intersite interaction, ternary and tricritical points approach the point of the continuous COI-NO transition at $\bar\mu=0D$. A special critical point of the ground-state phase diagram is when the ternary point, the tricritical point, and the point of transition at the half-filling merge together (see Fig. \ref{fig:GS-critical}a): $zV \approx 0.4D$ ($U=2zV$, $\bar\mu=0D$, $n=1$). Although we cannot distinguish the tricritical point from the ternary point even earlier due to their asymptotic approach to each other, it is clear that they can merge at $\bar\mu=0D$ only, because the COI phase ($n_\text{COI}=1$) cannot continuously transition to the NO phase at $\bar\mu\ne0D$ ($n\ne1$). Thus, for $zV \lesssim 0.4D$ the COM phase does not exist and the symmetry-breaking transition is always discontinuous except for the transition at $\bar\mu=0D$.

\section{Finite-temperature results}\label{sec:ft}

\begin{figure}
    \centering
    \includegraphics[width=0.99\textwidth]{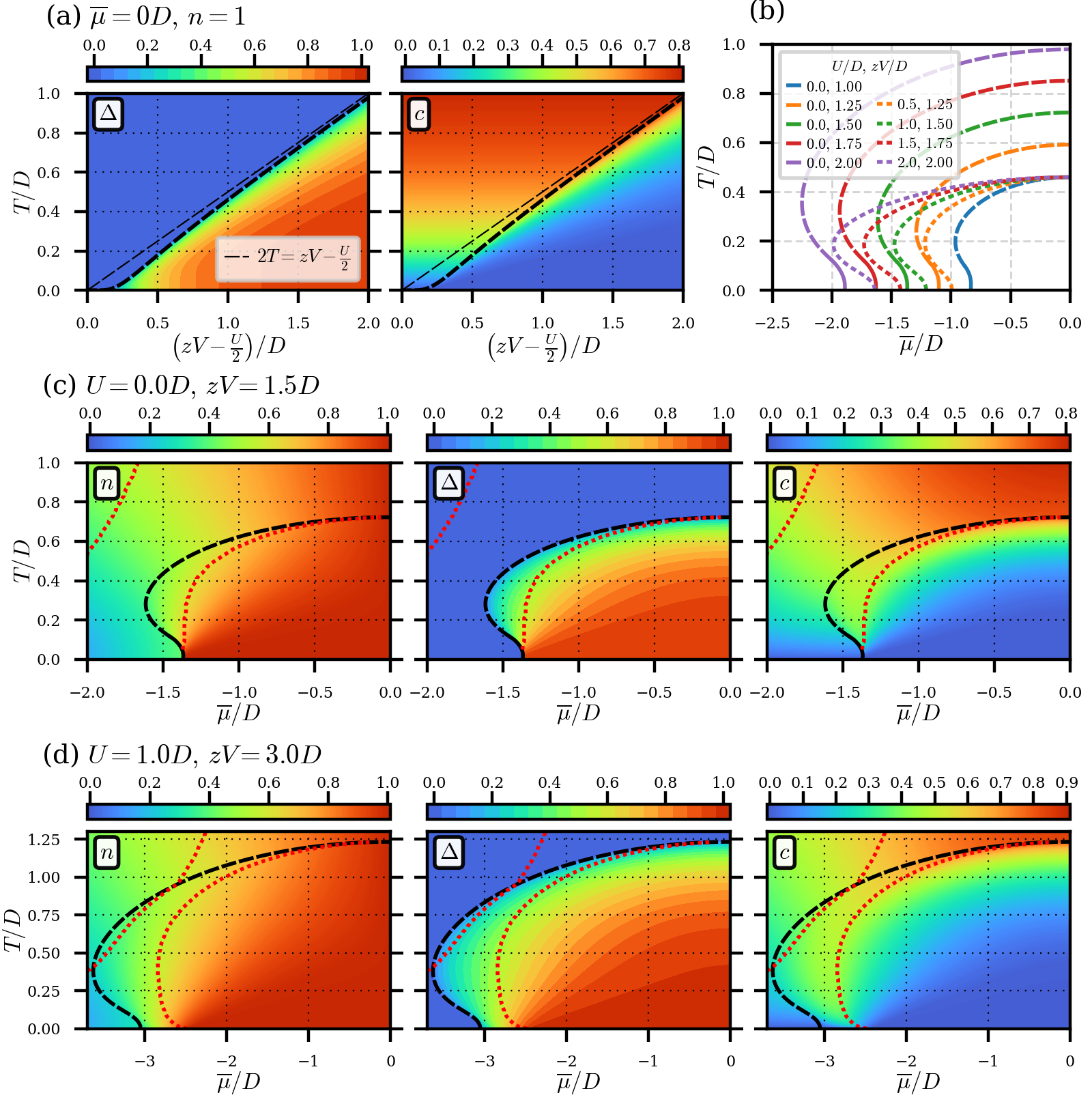}
    \caption{
    Finite-temperature phase diagrams. 
    (a) the $(zV-\frac{U}{2})$-$T$ phase diagram for $\bar\mu=0D$; (b)  the evolution of the $\bar\mu$-$T$ phase diagram when the interaction strengths change; 
    (c) and (d) the $\bar\mu$-$T$ phase diagrams for $U=0D$, $zV=1.5D$ and $U=1D$, $zV=3.0D$, respectively.
    On the phase diagrams: dashed and solid black lines are for continuous and discontinuous transitions, respectively; color scales are for concentration $n$, charge polarization $\Delta$ and thermally-exited charge carrier concentration $c$ (as labeled). Both short and long dashed lines on panel (b) denote continuous transitions and are different to distinguish the $U=0D$ lines and the $zV-\frac{U}{2}=1D$ lines. The red dotted lines on panels (c) and (d) show points at which the Fermi level is located on the edge of a band. Only the phases with the lowest grand potential are shown (no metastable phases).}
    \label{fig:finT}
\end{figure}

Finite temperatures destroy the charge order, i.e., in large enough temperatures, only the NO phase occurs. The character of this transition depends on the part of the phase diagram.

\subsection{Simplified expressions at $T>0$}\label{sec:ft-simexp}

For the $\bar\mu=0D$, there is one-to-one correspondence between the difference $(zV - \frac{U}{2})$ and a critical temperature $T_\text{c}$ of the continuous CO-NO transition (cf. also Refs. \cite{RobaszkiewiczPSSB1973,RobaszkiewiczAPPA1974}):
\begin{equation}\label{eq:finT-halffil}
    zV - \frac{U}{2} = \left(
    \frac{2}{\pi D^2} \int_{-D}^{D} d\varepsilon
    \frac{\sqrt{D^2 - \varepsilon^2}}{\varepsilon}
    \tanh\left[ \frac{\varepsilon}{2T_\text{c}} \right]
    \right)^{-1}.
\end{equation}
It is shown in Fig. \ref{fig:finT}a, and is also the highest temperature at which the charge order can exist (see phase diagrams in Fig. \ref{fig:finT}b). It has an asymptotic behaviour in the large $(zV - \frac{U}{2})$ limit:
\begin{equation}
    T_\text{c} \approx \frac{1}{2} \left(zV - \frac{U}{2}\right).
\end{equation}

For an arbitrary filling, the self-consistency equations for the continuous CO-NO transition (when it exists) are not simplified much, but it is still useful to find it solving
\begin{eqnarray}
    n & = & \int d\varepsilon \rho(\varepsilon) \left[
    f_\text{FD}(\frac{E_0 - \varepsilon}{T_\text{c}})
    + f_\text{FD}(\frac{E_0 + \varepsilon}{T_\text{c}}) \right],
    \\
    1 & = & \left(zV - \frac{U}{2}\right) \int d\varepsilon
    \frac{\rho(\varepsilon)}{\varepsilon}
    \left[
    f_\text{FD}(\frac{E_0 - \varepsilon}{T_\text{c}})
    - f_\text{FD}(\frac{E_0 + \varepsilon}{T_\text{c}}) \right],
\end{eqnarray}
instead of straightforward solution of system (\ref{eq:sceA})--(\ref{eq:sceB}) on a grid of parameters. Moreover, the same notes as
previously mentioned for the system of equations (\ref{eq:cont-COM-NO-A})--(\ref{eq:cont-COM-NO-B}) are valid: the convergence at the continuous transition point is extremely slow, and the solution with $\Delta \ne 0$ indicates the discontinuous character of the transition.

\clearpage

\begin{figure}
    \centering
    \includegraphics[width=0.9\textwidth]{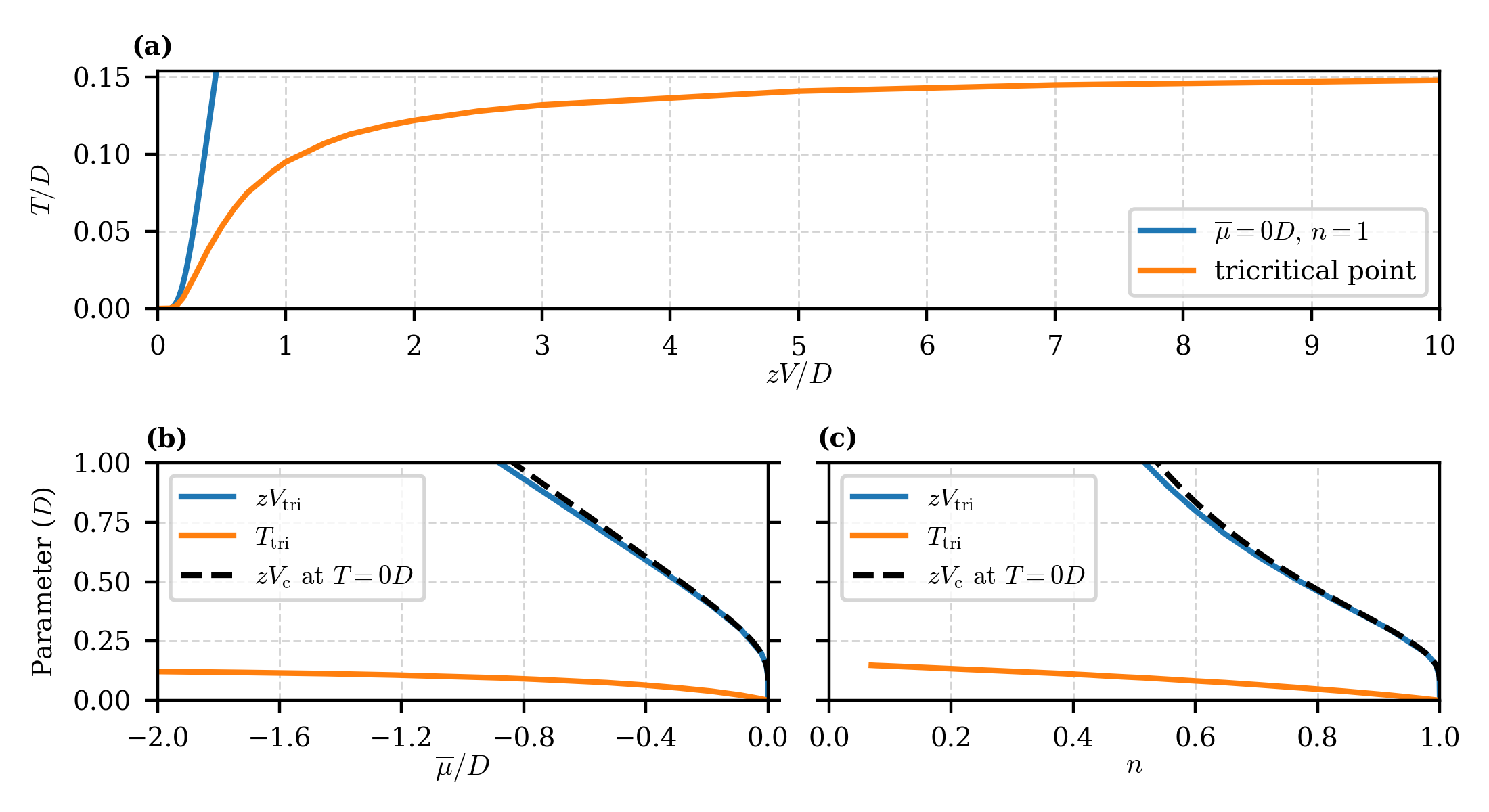}
    \caption{Tendencies of the finite-temperature tricritical point for $U=0D$: relation between the intersite interaction strength and temperature of the tricritical point, alongside with a point of the continuous CO-NO transition at $\bar\mu=0D$ (eq. (\ref{eq:finT-halffil}), Fig. \ref{fig:finT}a) (a); and dependencies of the intersite interaction strength and temperature of the tricritical point on the shifted chemical potential (b) and the electron density of the NO phase (c), alongside with a point of the discontinuous COI-NO transition at $T=0D$ ($V_\text{c}$, Fig. \ref{fig:GS}a).}
    \label{fig:finT-critical}
\end{figure}

\subsection{Results at $T>0$}\label{sec:ft-results}

An example of the  
finite-temperature phase diagram as a function of $\bar\mu$ for $U=0D$ and $zV=1.5D$ as well as for $U=1D$ and $zV=3.0D$ are shown in Figs. \ref{fig:finT}c and \ref{fig:finT}d, respectively.

At finite temperatures there is no clear distinction between the COI and COM phases.
The $U=0D$ phase diagram is similar to those presented in Refs. \cite{UV1993a,CRT2008,CGR2010,CGNR2012a} for the spinless fermion model.
At low temperatures, the CO-NO transition stays discontinuous, and thus there is a range of electron density that is possible within the phase-separated states only. On the scale of the shown plot (Fig. \ref{fig:finT}c) the metastability region of the CO phase inside the NO phase is not significant -- around $0.04D$ for $T=0D$ and smaller for larger $T$. 

With the increase of temperature the transition becomes continuous, thus, a tricritical point is identified 
($T_\text{tri}$, $V_\text{tri}$, $\bar\mu_\text{tri}$).
Solid and long dashed lines in Fig. \ref{fig:finT}b show the evolution of $U=0D$ phase diagrams as the intersite interaction $V$ varies. As it was clear from the combination of the $T=0D$ phase diagram (Fig. \ref{fig:GS}a) and the $\bar\mu=0D$ phase diagram (Fig. \ref{fig:finT}a), the parameter range of ordered phases proportionally increases when $V$ increases.
The coordinates of the tricritical point from the different perspective and for $U=0D$ are shown in Fig. \ref{fig:finT-critical}. The $zV$- and $T$-parameter ranges when the phase separation takes place is below the line of tricritical points in Fig. \ref{fig:finT-critical}a. The temperature in this region should not exceed $0.15D$ unless one deals with very large values of $zV$. 
We present data for the tricritical point for $U=0D$ only. It is not relevant for $U \gtrsim 0.3-0.5D$ at which there is no discontinuous CO-NO transition already for $T=0D$ (Fig. \ref{fig:GS-critical}a). When both nonzero $U$ (Fig. \ref{fig:GS-critical}a) and nonzero $T$ (Fig. \ref{fig:finT-critical}a) act together, the discontinuous transition and phase-separated states disappear even faster. Note that, for the model parameters when the tricritical point exists, the resulting structure of the $n$-$T$ diagrams including a region of the phase-separated state occurrence, is like the finite-temperature diagrams presented in, e.g., \cite{CRT2008,CGR2010,CGNR2012a}, at least qualitatively.

For $U>0D$, the COM phase appears at the ground state, but there is no well-defined COM-COI phase transition at finite temperatures. Some reminiscence of the transition are visible in the behavior of thermally-excited charge carrier concentration $c$ near the $T=0$ boundary (see also discussion at the end of the section).
The increase of $U$ reduces the temperatures in which the CO phase can occur. It follows from the fact that the maximum $T_\text{c}$ is at $\bar\mu=0D$ and depends solely on the difference $(zV-\frac{U}{2})$, as discussed above. The short dashed lines in Fig. \ref{fig:finT}b show the evolution of the phase diagram as the $(zV - \frac{U}{2})$ stays constant while both $zV$ and $U$ increase. The discontinuous CO-NO transition quickly disappears, and the whole region of the CO phases becomes elongated along the $\bar\mu$ axis.

A noticeable feature of the finite-temperature phase diagram is the reentrant behaviour of the CO phase: for a certain range of the chemical potentials, the CO phase is not stable at the ground state, but becomes stable as the temperature increases. Counterintuitively, around the entrance into this CO region (when moving from lower to higher $T$), the metallic CO phase has higher entropy, that comes from charge degrees of freedom, than metallic NO phase. Nevertheless, with further increase of temperature this situation changes, but the CO phase still stays stable (with the lowest $\Omega$). Hence, the reentrance behaviour itself can come from the competition between the electron itineracy ($t$-term) and localization due to the intersite repulsion ($V$-term) that favors the NO phase at the ground state, but the increase of temperature shifts the balance towards the CO phase.
Such a reentrancy region becomes larger when $zV$ increases (see Fig. \ref{fig:finT}b). Note that it is not only specific for the MFA used in this work, but such behavior occurs also for other approaches like the DMFT \cite{PBB1999,TSB2004}, what suggests that it is rather associated with non-local correlations. 

It is clear from the color map plot of the parameter $c$ (eq. (\ref{eq:c})) that the CO phase is rather insulating far from the CO-NO transition. Metallic properties is a necessary step towards the transition to the NO phase.
We also show the points at which the Fermi level is located on the edge of bands (the lower band in the case of the CO phase), i.e., where $A(i\eta)D$ becomes zero (see red dotted lines in Figs. \ref{fig:finT}c and \ref{fig:finT}d, cf. also Fig. \ref{fig:spectral}). For $\bar\mu$ larger than the right red line, the Fermi level is located inside the gap (cf. Fig. \ref{fig:spectral}b), however, the thermally excited charge carriers yield metallic properties for large enough $T$. Note that the right red line (located inside the CO phase occurrence region) also corresponds to the singularity (the lower one from $E_A$ or $E_B$ is equal to $0D$, i.e., it is at the Fermi level), hence the peak of the parameter $c$ is around this line for low temperatures (high temperature smears out this peak). For $\bar\mu$ smaller than the left red line, the Fermi level is below all bands of the system (cf. Figs. \ref{fig:spectral}a and \ref{fig:spectral}b for shape of spectral function in both phases). For small $zV$ the line is located only in the NO phase occurrence region.  It is interesting, that for large enough $zV$ (see Fig.~\ref{fig:finT}d) the CO phase can exist when the Fermi level is below all bands (i.e., the left red line goes also through the CO phase occurrence region).
This is possible because the concentration of thermally-excited charge carriers is enough to make their repulsion between sublattices (due to intersite $V$ interaction) in the NO phase disadvantageous.

\section{Conclusions and final remarks}\label{sec:conclusions}

The comprehensive analysis of the mean-field solution of the two-sublattice extended Hubbard model on the Bethe lattice is presented.
We have filled the gap in literature to identify the correlation-induced effects and interpret the non-strongly-correlated phase, and presented the results that can be used in pedagogical purposes.

Comparing our MFA results with the DMFT results \cite{kapcia2017} a few conclusions can be made. Most obvious, strong enough local correlations cause the formation of a non-charge-ordered Mott insulator around the $\bar\mu=0D$; and, for $U\gtrsim D$ and finite $zV$, the formation of a quarter-filled charge-ordered insulator on the border between NO and COM phases. Moreover, the strong-correlation effects remove the asymptotic behaviour of the COM and COI phases when approaching to the $\bar\mu=0D$ and $zV = \frac{U}{2}$. Within the DMFT, the continuous COI-NO transition at $\bar\mu=0D$ and $zV = \frac{U}{2}$ becomes discontinuous at around $zV = U$ which is the line that is easily predicted from the atomic-limit investigation, while the COM phase does not appear for $|\bar\mu| \lesssim \frac{U}{2}$ due to the Mott physics that has a strong effect closer to the half filling ($\bar\mu=0D$). Moreover, when the COM phase exists within the DMFT, the continuous COM-NO transition happens for smaller values of $zV$ than that obtained within the MFA, while, the line of the COI-COM transition is nicely reproduced by our mean-field results and less affected by the correlation effects. As a result, for $|\bar\mu| \gtrsim \frac{U}{2}$ the COM phase becomes wider when the Mott physics is taken into account as compared with our MFA results. The predicted with the DMFT half-filled COI phase above the line $zV = U$ (the same line as predicted from the atomic limit) is perfectly reproduced with our MFA results. The Mott physics is not manifested in this COI phase, and all ground-state formulas from Sec. \ref{sec:COI} are valid within the DMFT as well. Additionally, the same way the fully-occupied and fully-unoccupied phases are reproduced within the MFA, and hence, the borders $|\bar\mu| \ge zV + \frac{U}{2} + D$ can be used.

Worth noting, that the neglected here magnetic order would be especially relevant for $zV < \frac{U}{2}$.

We also showed the importance of analytical simplifications in the considered model. For the ground state, the numerical integration around sophisticated points (the continuous phase-transition points, the $\bar\mu=0D$ and $|\bar\mu|=zV + \frac{U}{2} + D$ points) leads to numerical inaccuracies that prevent from correct identification of the phase-transition lines. Moreover, the symmetry-broken phases require enormous number of iterations for the convergence close to the continuous transition to the non-charge-ordered phase, which can be avoided with analytical derivations. 
The mentioned problems appear when such analytical simplifications are not possible, and may be even harder to deal with when we have to sum over momentum vectors instead of the numerical integration. We expect that it is as well relevant for other self-consistent methods (such as DMFT) when one considers any ordering phenomena. We thus expect, that even in the limit of the infinite coordination number (where the DMFT is precise), it can be hard to extract the quantitatively precise results from DMFT for the continuous symmetry-breaking transitions.

One should mention that, for $U=0D$ the model  (\ref{eq:EHM}) is equivalent to a spinless fermion model \cite{VlamingJPCM1992,UV1993a,UhrigJPCM1993,ShankarRMP1994,UV1995,KK2001,ZTYT2005,FTZ2006,CRT2008,CGNR2012a,CGNR2012b}.
The structure of the diagram obtained in this regime is in agreement with the previous studies where the other shapes of the DOS were used (the qualitative differences for phase boundaries are very small) \cite{UV1993a,ShankarRMP1994,UV1995,KK2001,ZTYT2005,FTZ2006,CRT2008,CGNR2012a,CGNR2012b}.
Notice that for $U=0D$, the COM phase is not present on the phase diagrams and is unstable for any $n$ ($\partial n / \partial\mu < 0$).
However, the incommensurate orderings, which are not analyzed in this work, might also be possible with \cite{CGNR2012b} and without (when $zV$ is not too large) a next-nearest-neighbor hopping \cite{UV1993a,UV1995}.
In the presence of the next-nearest-neighbor hopping, the COM can be stable at the ground state \cite{CRT2008}.

On an exemplary model studied in this work, i.e., the extended Hubbard model,  we showed that, despite ignoring the correlation effects, the mean-field treatment of the model can provide significant insights into the problem (cf. \cite{ACH2010,kapcia2017}). 
The Hartree-Fock MFA overestimates the stability of long-range orders and associated critical temperatures, but it can, nevertheless, give a qualitative description of the system in the ground state in certain interaction parameter ranges. 
Moreover, the non-correlated phases are found within the dynamical mean-field theory when the intersite interaction prevails over the on-site interaction \cite{ACH2010,kapcia2017}, while some other phases and phase transitions can have similar qualitative features as within the MFA. For comparison between results obtained with exact and mean-field treatment of on-site Hubbard interaction and for various strongly correlated models see e.g., the extended Falicov-Kimball model \cite{LemanskiPRB2017,KapciaPRB2019,KapciaJPCM2021} and a model of superconductor with pair hopping \cite{KapciaAPPA2016,KapciaAPPA2018}).

\section*{Acknowledgements}

We thank Agnieszka Cichy for very fruitful discussions. 

\section*{Declaration of competing interest}

The authors declare that they have no known competing financial interests or personal relationships that could have appeared to influence the work reported in this paper.
The authors' institution had no role in the design of the study; in the collection, analyses, or interpretation of data; in the writing of the manuscript, or in the decision to publish the results. 

\section*{Data availability}

Data will be made available on request. 
All presented data are obtained by (numerical) solving the equations, which forms are included explicitly in the present work.

\section*{Funding sources}
This research did not receive any specific grant from funding agencies in the public, commercial, or not-for-profit sectors.

\printcredits

\bibliographystyle{elsarticle-num}

\end{document}